\documentclass[12pt, letterpaper]{article}

\usepackage{amsmath,amssymb}
\usepackage{cite}
\usepackage{fancyhdr}
\usepackage[top=1in, bottom=1in, left=1in, right=1in]{geometry}
\usepackage{graphicx}
\usepackage{hyperref}

\numberwithin{equation}{section}
\date{}

\title{{\rm\footnotesize \qquad \qquad \qquad \qquad \qquad \ \qquad \qquad \qquad \ \ \ \ \ \                  UTTG-31-14\ TCC-032-14     RUNHETC-2014-03    
SCIPP 14/02}\vskip.5in     Holographic Inflation Revised}
\author{Tom Banks\\
NHETC and Department of Physics \\
Rutgers University, Piscataway, NJ 08854-8019\\
{\it and}\\
Department of Physics and SCIPP\\
University of California, Santa Cruz, CA 95064\\
E-mail: \href{mailto:banks@physics.rutgers.edu}{banks@physics.rutgers.edu}
\\
\\
W. Fischler\\
Department of Physics and Texas Cosmology Center\\
University of Texas, Austin, TX 78712\\
E-mail: \href{mailto:fischler@physics.utexas.edu}{fischler@physics.utexas.edu}}
\begin{document}

\maketitle
\thispagestyle{fancy} 

\begin{abstract}
This paper is a major revision of our previous work on the HST model of inflation.  We identify the local fluctuations of the metric with fluctuations of the mass and angular momentum of black holes, and show that the consistency conditions in HST for a single trajectory to see more and more of a homogeneous distribution of black holes, imply that the system outside the horizon is undergoing inflation: small systems of equal entropy, are not in causal contact. Homogeneity then requires that the initial trajectory underwent inflation that expanded the black hole radius into our current horizon.  The low entropy of the initial state of the universe is explained by the fact that this is the maximal entropy state, which has long lived localized excitations, and which can form structures more complex than black holes.  The number of e-folds, reheat temperature of the universe and size of inflationary fluctuations are calculated in terms of a few parameters.
\normalsize \noindent  \end{abstract}


\newpage
\tableofcontents
\vspace{1cm}

\section{Introduction}

This paper is a major revision of our earlier work\cite{holoinflation12} on Holographic Space-Time (HST) models of inflation. In it we have incorporated the lessons we have learned about the origin of locality in our work on scattering theory and black holes in Minkowski space\cite{holounruh-fw3}.  As a consequence, we've had to revise some of our earlier estimates of parameters, but have obtained a major increase of clarity and coherence, and extended our model to include the reheating era, which was previously obscure in HST\footnote{Somewhat surprisingly, the reheating mechanism we will describe resembles elements of our earliest holographic models of cosmology\cite{holocosm1}}.   

Among the main new results in this paper we have

\begin{itemize}

\item The number of e-folds is fixed at the bound\cite{efoldbound} given by the present value of the cosmological constant (c.c.). More precisely, the range of scales over which our model predicts a nearly scale invariant fluctuation spectrum ranges from the size of the cosmological horizon, $N L_P$ to the size of the inflationary horizon $n L_P$ multiplied by the ratio of the FRW scale factors $\frac{a_{NOW}}{a_I}$ between the present, and the end of inflation.

\item The reheat temperature of the universe is fixed in terms of the inflationary Hubble scale, the size of primordial scalar fluctuations and the density of black holes in an early phase of the universe, which is dominated by a dilute black hole gas.
For typical high scale values of these parameters we have $T_{RH} \sim 10^7 - 10^8 $ GeV.  

\item The formula for the tensor to scalar ratio $r$ has an extra factor of $\epsilon \equiv - \frac{\dot{H}}{H^2} $, compared to the corresponding formula $ r = 16 \epsilon $ for QUEFT\footnote{The acronym QUEFT stands for Quantum Effective Field Theory.  Jacobson's Principle\cite{ted} tells us that classical gravitational field equations are always valid as a hydrodynamic description of a theory of quantum gravity obeying the Covariant Entropy Principle.  These equations should be quantized only when studying small excitations of a system with a ground state, like asymptotically flat or AdS space-times.  A central contention of our work is that QUEFT is not a valid approximation during the inflationary era.} inflation.  The explanation of this factor is simple.  In HST, the curvature fluctuations $\frac{\delta H}{H}$ are of order $\frac{H}{m_P}$ while in QUEFT models they are smaller by a factor $\sqrt{\epsilon}$.  The numerical significance of this remark is mitigated both by the fact that in HST $N_e \sim 80$ so that $\epsilon$ is not too small, and the fact that we do not yet know how to calculate the ``order one" coefficient $s$ in the formula $r = s \epsilon^2 $, from the HST model.

\item The approximate $SO(1,4)$ symmetry postulated in our previous work is identified explicitly and connected to the $SL(2)$ symmetry identified in our model of the pre-inflationary $p = \rho$ era.  During the inflationary era, the Hamiltonian of our system is a generator $J_{04}$ of $SO(1,4)$, identified in each horizon volume with the $L_0$ generator of $SL(2)$.
The $SO(3)$ subgroup, which commutes with $J_{04}$ is an exact symmetry of the formalism, and we argue that, after a few e-folds of HST-inflation, the density matrix of the system is invariant under this $R \times SO(3)$ subgroup and approximately under the full dS group.

\item Once they come into the horizon, individual horizon volumes behave like black holes and their fluctuations are fluctuations of the mass and angular momentum of these black holes, which contribute to scalar and tensor fluctuations respectively.  The sizes of these curvature fluctuations are comparable and their two point functions are approximately $SO(1,4)$ invariant.  The conventional scalar fluctuation in the metric is related to the scalar curvature fluctuation by an inverse factor of the slow roll parameter 
$\epsilon = \frac{\dot{H}}{H^2} $.  This accounts both for its dominance over the tensor fluctuations and for the deviation from scale invariance of its spectrum. 
Differences between predictions of HST and QUEFT for the scalar two point function can be absorbed into the choice of the background slow roll function $H(t)$.

 The tensor tilt, by contrast, is zero in HST.  An additional contribution to the primordial gravitational wave spectrum comes from decay of the black holes.  It is suppressed by $\frac{1}{g}$, the number of effective massless particles into which the holes decay, but might dominate over the quantum tensor curvature fluctuations if the slow roll parameter is small enough. Its spatial profile mirrors that of the scalar fluctuations, and so would have a red tilt equal to that of the scalar fluctuations.  Either of these contributions are distinctly different from the predictions of QUEFT models (at least single field models) and a measurement of the tensor two point function over a sufficiently large range of wave numbers could differentiate between the two models.  
 
 \item  In both QUEFT and HST models, the combination of Maldacena's squeezed limit theorem and approximate $SO(1,4)$ invariance implies that all three point functions containing a scalar curvature fluctuation are smaller than the scalar two point function by a factor $ (n_S - 1) \frac{H}{m_P} \sim 10^{-7} $.  
The two kinds of model make distinctive predictions for the tensor three point function, but the measurement of that quantity currently lies in the realm of science fiction, because the rumors from the Planck and BICEP2 collaborations suggest that we will soon have a firm observational bound of about  $ r < 0.1$.

\end{itemize}

One of the novel features of our current presentation of HST cosmology is the necessity to distinguish, for the first time, between the time parameter of our quantum model and the usual FRW time-slicing.  In the cosmology of the very early universe, there are no localized excitations.  It was not necessary, and physically impossible, to distinguish between what was occurring at one point of our time slicing, and another.  Indeed, in the $p = \rho\rightarrow p = -\rho$ cosmology, called Everlasting Holographic Inflation (EHI), which we will review in section 3, all the DOF should be thought of as living on the apparent horizon of the the causal diamond of some time-like trajectory originating on the Big Bang hypersurface.  Their Hamiltonian is a fast scrambler\cite{Susskind-Sekino} and does not behave like a local field theory on the holographic screen. None of the constraints, which define bulk localized excitations in HST, are imposed on the state of the system.  As a consequence we were free to identify our quantum Hamiltonian time with FRW time, since, as one can see in Figure 1, the two definitions of time coincide at the position of the trajectory on the corresponding time slice.  We have made a number of errors in previous presentations of HST inflation, by trying to continue making this identification.  Instead, as one can see by referring to Fig. 1, the time slices in our QM refer to hyperbolae which interpolate between two null cones with a small proper time difference between their tips.  Far away events on these time slices, are seen as they were a long FRW time in the past.  This distinction will be important in section 4.

\begin{figure}[htb]
\begin{center}
  \includegraphics[width=0.80\textwidth]{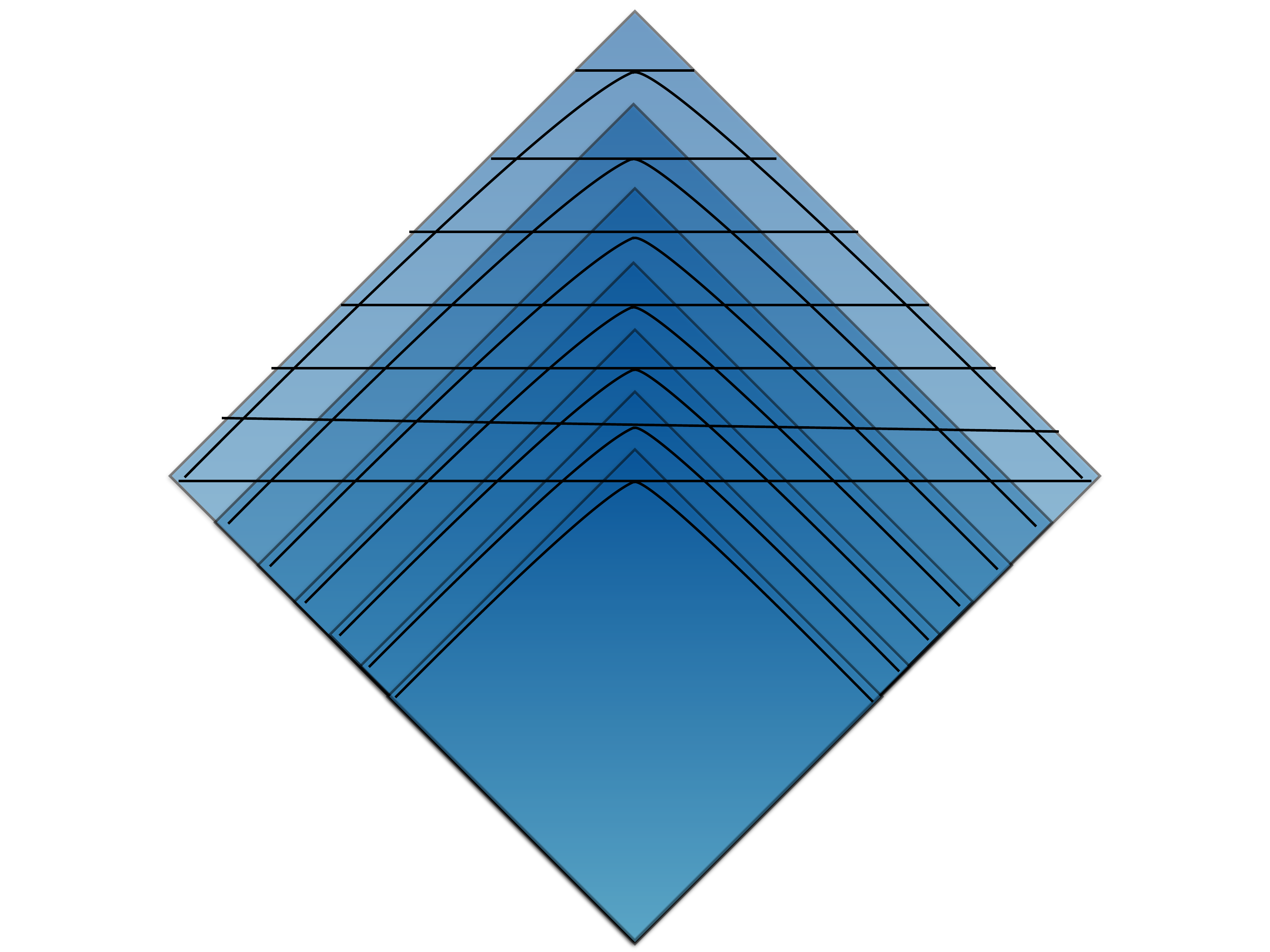}
   \caption{Horizontal slices are FRW, hyperbolic slices are HST.\label{fig1}}
   \end{center}
\end{figure}

Throughout this paper, we will be utilizing the prescient view of general relativity invented by Ted Jacobson\cite{ted} and expanded upon by others\cite{paddyverlinde}.  Fields and field equations\footnote{String theory has taught us that we should view all effective fields as originating from supergravity and geometry in higher dimensions, so Jacobson's insight should be generalized to fields that are not explicitly associated with geometry in four dimensions.} are the hydrodynamic equations of quantum systems obeying the BHGFSB\cite{BHFSB} entropy/area relation between the Hilbert space associated with a causal diamond and the area of that diamond's holographic screen.  These equations, like all hydrodynamic equations, should be quantized only when studying low lying bulk localized excitations of quantum gravitational systems with a unique ground state.  String theory in asymptotically flat and AdS space-times {\it does} have a unique ground state, and this accounts for the success of effective field theory methods in that context.  As we'll see, quantized effective field theory is misleading in the very early universe.
Classical field equations on the other hand, are valid for the quantum system  in high entropy states where hydrodynamic averages are useful coarse grained descriptions.  In our attempt to model the universe, we will describe quantum models, and then fit Lorentzian geometries to their hydrodynamics.

In Section 2. we will give a brief review of our work on Minkowski and dS space-times.  Notable features will be the identification of horizon entropy with low energy DOF not captured by the local excitations of quantum field theory.  Localized states are in fact constrained states of these DOF, and their energy is defined in terms of the constraints, and conserved asymptotically because it enforces an infinite number of constraints on an infinite causal diamond.  The HST model has a clear description of both particle and black hole states, and the transitions between them.
Section 3. is a review of the EHI cosmology and its approximate $SL(2)$ symmetry.  The FRW description of this system is a good description in the limit of large causal diamonds and the real system has no singularity.
Section 4. is the core of this paper.  It describes the inflationary model, which we believe is relevant to the universe we observe.  We derive bounds on the maximum temperature of the universe, which are related to the values of inflationary parameters.  This model makes it very explicit that one {\it must } choose low entropy initial conditions in order to have local excitations in the universe.  Further constraints come from insisting that the local excitations are more complex than a few large black holes or the radiation from their decay.  We call this excuse for the low entropy initial conditions a {\it topik\`{e}s-thropic} explanation, from the Greek word {\it topik\`{e}s} for {\it local}.  We show that more refined versions of this argument put an upper bound on the reheat temperature of the universe in the HST model in terms of parameters characterizing the inflationary era.  We also argue that in this framework the number of e-folds is essentially given by an upper bound we announced some time ago\cite{efoldboundseealsoalbrecht}.
In this section we also give a brief review of observational signatures of this model.  A more comprehensive paper about the predictions for two and three point functions of fluctuations will appear shortly\cite{tbwftj}. 

Section 4. also contains brief comments about baryogenesis in the HST model.  Our bound on the reheat temperature allows many conventional low energy mechanisms for baryogenesis, but rules out high scale leptogenesis. We also point out the possibility of producing the baryon asymmetry during the era of black hole decay by applying anthropic arguments to the statistics of black hole decay.  Thermodynamic averages of the decay products are symmetric, but this allows for the possibility of unlikely histories, which produce an anthropically selected asymmetry if the model has no microscopic mechanism for baryogenesis.  Anthropic arguments in HST cosmology are complicated, and differ significantly from model independent arguments of this type, or those based on the string landscape.  They will be covered in future work.

Section 5. is our conclusion, and recapitulates the basic structure of the model in a way which uses a minimum of matrix model formalism.  The basic idea of the model of this paper is very simple.  If we take the HST point of view that physics is done separately along different time-like trajectories, and that most of the degrees of freedom in a finite causal diamond are inaccessible to local near-geodesic observers inside the diamond, then a universe containing localized excitations must have finely tuned initial conditions.  The minimal fine tuning for a given amount of matter\footnote{The phrase ``amount of matter" seems ill defined, but in asymptotically dS space time, with asymptotically large Hubble radius, it has a precise definition in terms of the entropy deficit of a state with localized excitations.} has the matter appear as a dilute gas of black holes.  A universe with a phase dominated by a normal gas of matter or radiation will only evolve from a fairly uniform black hole gas.  Absolute uniformity is impossible since the individual black holes are chaotic quantum systems with finite entropy, $S$.  There are fluctuations of their mass and angular momentum of order $1/\sqrt{S}$, and these are fluctuations in the scalar and tensor components of the Weyl tensor.  Reheating of the universe comes from black hole decay, and the reheat temperature is related to the entropy of the individual black holes, and thus to the size of inflationary fluctuations.

The fact that black holes that enter the horizon just before the onset of c.c. domination of our universe, have to be isolated quantum systems with entropy $S$, implies that the universe had to undergo inflation up to a conformal time halfway between the Big Bang and the ultimate dS horizon.  The black holes that come into the apparent horizon of a trajectory, as it expands, are composed of the thermalized horizon degrees of freedom of individual horizon volumes of a slow roll cosmology.  In field theory models of inflation, all of these DOF are frozen into a unique adiabatic ground state, whereas in HST they are independent thermalized systems, which wander chaotically through a large Hilbert space of states.

\section{Minkowski/dS Space}

In our opinion, the theory of scattering in quantum gravity should be based on representation theory of the generalized super-BMS algebras (GSBMS)\cite{agstb}, on the conformal boundary of Minkowski space. This is discussed in the Appendix.   The generators of this algebra are spinors $\psi_{\alpha} (P, p)$, where $P^2 = 0$ is an in or outgoing momentum at null infinity, and $p$ labels the spectrum of the Dirac operator on compact internal dimensions, with an eigenvalue cutoff. When we retreat from the conformal boundary to a causal diamond with a finite area holographic screen, we use the Holographic Principle/Covariant Entropy Conjecture\cite{BHFSB} as our guide.  The Hilbert space which is the smallest representation of the appropriate deformation of the GSBMS algebra, must be finite dimensional.    The obvious deformation, the only one that preserves rotation invariance, is to restrict $\psi_{\alpha} (P,p)$ to sections of the spinor bundle on the sphere, which are sums of Dirac eigen-sections with a bound on the eigenvalue.  This is the same as an angular momentum cutoff and the resulting variables are sections of the spinor bundle over the fuzzy sphere.  

Thus, in a finite causal diamond, of size $n$, the GSBMS generators become $n \times n + 1$ matrices $\psi_i^A (p)$ with anti-commutation relations
\begin{equation} [\psi_i^A (p) , \psi^{\dagger\ j}_B (q) ]_+ = \delta_i^j \delta^A_B Z(p,q) . \end{equation}
This is supplemented by commutation relations for the ``wrapped brane charges" $Z(p,q)$ with themselves and the fermionic generators, that close on a finite dimensional superalgebra.  The smallest representation of this super-algebra, for fixed values of $i,j,A,B$, must be
unitary, finite dimensional, and generated by the action of the fermionic generators on a single state, and is called the {\it pixel Hilbert space}. The $n \times n$ matrices $M(p,q)^j_i = \psi^{\dagger\ j}_A (p)\psi_i^A (q) $ will be the objects from which we construct the Hamiltonian.

At null infinity, the generators $\psi (P,p)$ are constrained to vanish at finite $P$ unless $P$ is contained in a finite set of spherical caps of finite opening angle on the sphere.  These define jets of particles.  The $P \rightarrow 0$ generators are allowed to be non-vanishing everywhere, except in annuli surrounding those caps.  They represent the possibility that soft particles are emitted or absorbed, with vanishing energy in any finite solid angle, but non-zero total energy through the sphere at infinity.  Representations of the algebra satisfying these constraints are called {\it jet state representations}. The finite diamond version of the jet state constraint is the requirement that $n K + Q$ of the matrix elements of all the $M_n (p,q)$ vanish when applied to a jet state, with $K,Q \ll n$.  The subscript on $M_n$ denotes the fact that they are built from the $n \times n + 1$ $\psi_i^A$ variables. We can arrange the non-vanishing matrix elements as
\begin{equation} \begin{pmatrix} K_1 & 0 &0&\ldots & 0 \cr
                                 0& K_2 & 0& \ldots & 0 \cr
                                 \vdots & \vdots & \vdots & \vdots & \vdots \cr
                                 0 & 0 & 0 & \ldots & n - K \cr \end{pmatrix}, \end{equation}
where $K = \sum K_i $.  The large block represents the zero momentum particle states in the $n \rightarrow\infty$ limit, while the smaller blocks are the energy bearing jets.  Most of the area of the screen is covered by ``zero" energy stuff, while energy emerges from a finite number of cones of small opening angle.
Correspondingly, we write a Hamiltonian
\begin{equation} H_{in} (n) = \sum K_i + \frac{1}{n^2} {\rm Tr}\ P(M_n) . \end{equation}
't Hooft scaling guarantees that the second term vanishes as $n \rightarrow\infty $, while the first sum can be written as the sum of bilinears in the fermions which converges to the momentum as the $K_i \rightarrow \infty$ at fixed ratio and with $K/n\rightarrow 0$.  The polynomial $P$ is of finite, $n$ independent, order. It will be constrained by the requirement that scattering amplitudes be Lorentz invariant, but we do not yet know how to implement those constraints. 

Note that the second term in $H_{in} (n)$ is invariant under $U(n) \otimes U(n + 1)$ acting on the variables $\psi_i^A (p)$.  In the limit, this converges to a group containing the group of area preserving mappings on the sphere.  The terms involving particles will be chosen to be localized in cones of finite opening angle, and break this invariance.  Hamiltonians with area preserving invariance do not preserve the local differentiable structure of the sphere and will not behave like local field theory.  Instead, following Sekino and Susskind\cite{Susskind-Sekino}, we conjecture that the Hamiltonian is a fast scrambler.  For our purposes it is not necessary to believe the strong conjecture that all such Hamiltonians are fast scramblers.  We need only a sufficiently rich choice of polynomials to have this property, that we are able to also impose the constraints of Lorentz invariant scattering amplitudes.

The subtlety of the correlated limits above suggests the wisdom of doing everything in an extremely large but finite causal diamond of size $N$.  The Hamiltonian $H_{out} (n)$ acts on the matrix elements of $\psi_i^A $ with $n > i \leq N$ and $n + 1 > A \leq N + 1$.  Its form, and the choice of initial state in this tensor factor of the Hilbert space are constrained by consistency with the Hamiltonians $H_{in} (n , {\bf x})$ of other trajectories.  At time $-N$ we consider an initial state satisfying the jet constraint above (with $n \rightarrow N$).  At time $-n$ the variables corresponding to some of the small blocks will be in $H_{in} (n) \equiv H_{in} (n, {\bf 0}) $.  Those that are not, will be in $H_{in} (n , {\bf x})$ for some other values of ${\bf x}$.  The labels ${\bf x}$ label a sampling of trajectories, rich enough to constrain the dynamics completely.  They form a topological lattice with the topology of a Cauchy surface, which we will always take to be that of flat three space.
Consistency between the two quantum systems at different values of ${\bf x}$ is achieved by putting a copy of $H_{in} (n, {\bf x})$, with appropriate modification of the number, sizes, and angular positions of the small blocks, into $H_{out} (n, {\bf 0})$.  This strategy is successful as long as the positions are sufficiently far away from each other in the geometry defined by causal relations. As a consequence, the system has, roughly, a local structure similar to that of a quantum field theory.  One can argue\cite{fw3} that this leads to a computation of many amplitudes in terms of time ordered Feynman-like diagrams.

Those amplitudes are ones in which the outgoing state of a causal diamond of size $n$ satisfies of order $En$ constraints with $E \ll n$, so that it can be interpreted in terms of jets of particles.  The factor of $1/n^2$ in front of the Hamiltonian, combined with 't Hooft scaling, insures that this is likely when $n$ is large.  $H_{in} (n)$ controls the evolution of degrees of freedom inside the causal diamond between the times $-n$ and $-n + 1$, as well as $n - 1$ and $n$.  Since the part of the Hamiltonian that can change block size has a small operator norm, particles will, to a large extent, retain their identity.   In some smaller nested causal diamond, this will no longer be true, but the outgoing state may still be constrained and be interpretable as a new set of particles. Such an amplitude may be visualized as an $m \rightarrow p$ particle vertex, localized in the small diamond in which the blocks changed size and angular orientation.  The consistency conditions then allow us to string together these vertices into Feynman diagrams.  

On the other hand, it may be that the final state has no constraints on its variables, in some fairly large diamond of size $R_S$.  The fast scrambler nature of the Hamiltonian, will bring the system to equilibrium.  This is in fact the inevitable result when the number of constraints $En$, which we would like to impose to define a particle state, is of order the total number of variables $R_S^2$. That is what happens, precisely when 
$E \sim R_S$.  The initial state is then very special, and the Hamiltonian will act to change it into a more generic state.  We've thus derived the parametric form of the formula for the Schwarzschild radius as a function of energy.   One can also show that Newton's law has the right scaling as a function of energy and impact parameter, and that, with the appropriate generalization of the model, all of this works in any space-time dimension\cite{holonewton}.

If we imagine following up on the creation of this equilibrium state in a larger causal diamond, it will behave like a localized object, since the constraints on the $E (n - R_S)$ variables in that diamond are not removed by the creation of the ``black hole".  Assuming that the trajectory we are following is at rest in a frame not too different than the center of mass frame of the collision\footnote{We are being a little disingenuous here.  We have not discussed, and do not have a full understanding of, spatial momentum conservation in these models.} then this object will move slowly w.r.t. the trajectory.   It indeed has all of the characteristics we expect of a black hole.   The outgoing matrix state will be of block diagonal form, but the small block of variables, of size $R_S$, will not be in the pure state that would characterize a particle of that energy, but in a maximally uncertain state.   The probability that those variables could be in a state where the $R_S \times R_S$ matrices were further block diagonalized into
$\epsilon \times \epsilon \oplus R_S \times R_S$, with $ \epsilon \ll R_S$, and the $\epsilon$ block in a time independent pure state is of order $e^{ -\epsilon R_S}$, since that is how many constraints would have to arise spontaneously.  This is a thermal distribution of particle energies, with temperature $\sim 1/ R_S$.  

We want to emphasize that these properties are quite universal in the class of models we study, as long as the order of the polynomial $P$ is $\geq 7$ (this bound is necessary to get Newtonian scaling of large impact parameter scattering \cite{holonewton}).  Most of these models are not acceptable theories of quantum gravity, because they won't satisfy the HST consistency conditions for trajectories with arbitrary relative velocity, which imply in particular that in the correlated $K_i , N \rightarrow \infty$ limit, scattering amplitudes become Lorentz invariant.  Experience from string theory shows us that we can expect these constraints to be pretty strong, and most particle spectra will not be consistent with them.  What's remarkable is that many qualitative features of quantum gravity are contained in {\it all } of these models, so that we can address many questions without solving all the constraints.

We'd like to end this section with a brief comment about the theory of stable dS space\cite{dS}.  In the limit of large dS radius, we propose that it consists of keeping $N$ finite, and letting the system evolve forever with the Hamiltonian $H_{in} (N)$.  This describes a single horizon volume of the global dS manifold.
The system obviously evolves to an equilibrium state, with infinite temperature, if we start from a generic state of the large block in the block diagonal form of the constrained initial state .  All constraints are erased in a time of order $N {\rm ln}\ N$, since the characteristic energy scale of the Hamiltonian is $1/N$ and the system is a fast scrambler\footnote{This estimate does not take into account other conservation laws, like electric charge, and the possibility of finely tuned initial conditions.  An electron, at rest at the origin of the static coordinate system of the causal diamond will have a decay time exponentially longer than this, of order $N {\rm ln}\ N e^{ m_e N} $. }.   Localized excitations which were not put in as a part of the initial conditions, will, after the scrambling time, appear only as short lived thermal fluctuations.  The probability of such an excitation, with energy $E$ is $e^{- EN}$, which is thermal at the dS temperature.  This is because what we call $E$ is a count of the number of constraints that define a particle state.   This explanation of the dS temperature as a result of a relation between energy and constraints in a maximally uncertain system, is evident in the formula for black hole entropy in dS space.  The presence of a black hole decreases the entropy of dS space, and when its Schwarzschild radius is $\ll $ the dS radius, that decrease is precisely a Boltzmann factor.

\section{The $p=\rho\rightarrow p= - \rho$ Model: E(verlasting) H(olographic) I(nflation)}

The HST construction of a space-time contains an infinite set of quantum systems, describing physics as seen from an infinite collection of time-like trajectories.  We will be describing a Big Bang universe, which ends in a dS final state, with Hubble radius given by $\pi (RM_P)^2 = L N (N + 1). $.  It will be an FRW model and we may take the topology of the initial value surface at the Big Bang to be that of flat space.  As in the discussion of Minkowski space, we introduce a sampling of trajectories defined by a lattice on that hypersurface.  Only the topology of that lattice is relevant, since geometry will be defined by causal structure, which is incorporated in the quantum dynamics. It is convenient to think about a cubic lattice, and label its points by a discrete three vector ${\bf x}$.  

We will be describing an FRW universe and each trajectory will have the same quantum system attached to it.  There will be no fine tuning of initial conditions of each individual quantum system required, in order to impose homogeneity and isotropy.  Homogeneity and isotropy make it easy to impose the fundamental consistency conditions of HST, namely that the density matrices describing shared information along any two different trajectories, are unitarily equivalent to each other.  We think it's likely that, given the chaotic dynamics we will prescribe for individual trajectories, no inhomogeneous solutions to the consistency conditions exist.  In the next section we will introduce models that have local excitations, and find plenty of inhomogeneous and consistent models.

The prescription along an individual trajectory, is that the Hamiltonian at time $t$ Planck units from the Big Bang can be written

\begin{equation} H (t) = {\rm Tr}\ P( M_n ; t)  + H_{out} (t) . \end{equation} In the EHI model, it will turn out that we do not need to specify $H_{out}$, except that it is constructed, at each time, from variables that do not appear in $H_{in}$.  The first part of the Hamiltonian, called $H_{in} (t)$ depends only on the bilinear matrices $M_n (p,q)$ constructed from the spinor variables $\psi_i^A (p)$, with indices running over $n (n + 1)$ values. It is the trace of a polynomial of some fixed order, but with coefficients that are chosen randomly at each time.  Susskind and Sekino\cite{Susskind-Sekino} conjecture that even with fixed coefficients such a Hamiltonian is a fast scrambler.   It randomizes a perturbation applied to a small number of variables in a time ${\rm ln} n t_0$, where $t_0$ is the characteristic time scale of the Hamiltonian (the inverse of the operator bound, if the eigenvalues are distributed relatively uniformly).  When the coefficients vary randomly with time, the results of \cite{haydenstanford} make it almost inevitable that the Hamiltonian will be a fast scrambler.  

The overlap rules are defined as follows: if two points are separated by $d$ lattice steps, then at time $t$ the overlap Hilbert space is generated by the sub-algebra of operators $\psi_i^A$, which were acted on by $H_{in} (t - d)$.  Our assumption that the initial state on the full Hilbert space, and sequence of time dependent Hamiltonians, are identical for all lattice points, guarantees that the density matrix of this subalgebra of operators at time $t$ is identical for both systems.  Homogeneity and isotropy\footnote{Isotropy is built into the model by the trace form of the action, which is $SU(2)$ invariant.  It also appears in the overlap rules because the overlap condition depends only on the number of lattice steps between points.  As $d$ gets large the locus of all points $d$ steps away from a given one becomes dense on a tilted hypercube.  In the distance defined by causality, the geometry of this hyper-cube is spherical. Since we view the lattice as only a sampling of the possible trajectories in space-time, this is evidence for exact isotropy in the overlap rules, which is certainly what one sees for the quantum mechanics of each individual trajectory.} reduce this infinite set of conditions to a rather simple one.  

We insist finally that, as $t\rightarrow\infty$ the Hamiltonian approach that of a $1 + 1$ CFT, with central charge $\sim t^2$.  The UV cutoff  on the CFT is proportional to $t^{-1}$ and the length of the half line on which it lives proportional to $L t$.  We assume $L \gg 1$ so that the one dimensional volume is large in cutoff units and the CFT description makes sense.  In a generic state, the expectation value of the energy is thus $ L t $, while the entropy is $ L t^2$.  The volume of the region inside the horizon, in the emergent geometry defined by our causal rules is $L^{3/2} t^3$.  Thus, the energy density, entropy density, and time are related by \begin{equation} \rho \sim L^{ - 1/2 } t^{-2} = L^{1/2} \sigma^2 .\end{equation}   These are the Friedmann equation, and equation of state of an FRW geometry with $p = \rho$.  We see Jacobson's principle in action: the hydrodynamic relations of a quantum system obeying the area law are equivalent to Einstein's classical field equations.  The quantum excitations of the system have nothing to do with the quantized Einstein equations.  Notice also that the singularity of the classical cosmology is a fake.  The emergent geometry is an approximate notion, valid only for large $t$, while the singularity occurs near $t = 0$, where the actual quantum system has a perfectly finite description.  Note that there is no curvature term in the Friedman equation, a fact which is linked to our ansatz of approximate scale invariance for the large $t$ Hamiltonian.  This is a property of the model, rather than a fine tuning of the choice of initial state.  In the $p = \rho$ model, homogeneity, isotropy and flatness are consequences of the choice of Hamiltonian, plus a random choice of initial state.  The model has no local excitations.

There is a simple modification of this model in which we declare the full Hilbert space at each site to have finite dimension $n \gg 1$.  We then propagate the system forever with the Hamiltonian $H_{in} (n)$, which is approximately that of a CFT.  We make an asymptotic change of time coordinate by going to a conformal frame where $< H > \sim 1/n$ (this is the analog of the change between FRW and asymptotic static coordinate times for an asymptotically dS FRW space-time).  We modify the overlap rules so that points that are more than $n$ steps apart on the lattice have no overlap Hilbert space, no matter how long a time has passed.. The asymptotic model on a single site is the same as that we used for a single horizon volume of dS space in the previous section.  However, we now have many copies of that volume, which don't communicate with each other.  This model, called the EHI model, depicts a transition from a non-singular Big Bang, with Planck scale horizon radius, through an approximately scale invariant $p = \rho$ phase, to a de Sitter phase with horizon radius $n L_P$.  The coarse grained hydrodynamics of this model is described by the FRW scale factor $a(t) = \sinh^{1/3} (3t/n) $, with time measured in Planck units.  The only local excitations in this universe are thermal fluctuations at late time.  Although the Hilbert space contains states\footnote{We're assuming here that the commutation relations of the $\psi_i^A (p)$ produce a particle spectrum like that in our universe.} which resemble the current, pre-de Sitter phase of the universe we inhabit, they are so improbable that not even anthropic arguments can redeem them.

In the next section we will build a model which does produce local excitations, by fine tuning initial conditions. We will see that that model must undergo an inflationary phase in which it resembles a ``coarsely woven lattice" of trajectories in the EHI model, consisting of trajectories which have no overlap whatsoever in Everlasting Inflation.  These are separated by $n$ lattice steps on the original lattice.  In the Holographic inflation model, these disjoint horizon volumes are put together into a single Hilbert space as the horizon expands.

\section{Inflation in HST}

The model of the previous section provides the clue for constructing a model of inflation.  We want to construct a model, which evolves in the future, to a dS space of radius $N \gg n \gg 1$, but ends up for a long period in a state with many meta-stable localized excitations.  The entropy formula for the Schwarzschild de Sitter black hole tells us that this is a low entropy state, and our matrix model represents it as a state in which the matrices $M_N (p,q) $ are block diagonal.  Although the mathematical structures are quite different from field theory inflation models, our model will identify regions of the sky of a given time-like trajectory, which start their sojourn inside the horizon as independent fluctuating quantum systems, with a high degree of bulk localizability.  They will all have the same number of DOF, the same Hamiltonian and the same initial state, and one can think of them as ``originating" from different points of the ``coarsely woven lattice" of the previous section.  However, the Hamiltonian is a fast scrambler and these systems will become coupled to $H_{in} (t)$ of the trajectory, at a rate which is unrelated to their internal clock.  Thus, in the view of that trajectory there will be fluctuations of the states of those systems, when they come into the horizon.  We will characterize those fluctuations below.

The matrix model of the post inflationary era in which the horizon expands, when compared with the HST theory of Minkowski space, will identify these systems as black holes, and the fluctuations as fluctuations in the black hole mass and angular momentum.  Thus, there are fluctuations of both the spin zero and spin two parts of the Weyl tensor.  Since these fluctuations are those of a statistical system with large entropy $n^2$, they are nominally Gaussian, and of the same order of magnitude, $ \frac{1}{n}$. The corrections to Gaussian statistics are down by at least another power of $n^{-1}$.  Recall that the conventional gauge invariant measure of scalar fluctuations, $\zeta$ is, in comoving gauge\cite{lythliddlebardeen?}
\begin{equation} \zeta = \frac{H^2}{\dot{H}} S = \frac{S}{\epsilon}, \end{equation}
where $S$ is the scalar curvature fluctuation.  Thus
\begin{equation} \zeta = \frac{\delta\rho}{\rho} \sim \frac{1}{n\epsilon} \sim 10^{-5} . \end{equation}
\noindent The last equation incorporates the observations on the CMB.  We can take the absence of tensor modes in the latest CMB data to be evidence that
$\epsilon$ is small.  Note that for small $\epsilon$ this is larger than the estimate of $\zeta$ in slow roll inflation models.  In those models, $\delta H / H\sim V^{\prime} / V \delta\phi \sim \sqrt{\epsilon} \delta\phi, $ and $\delta\phi$ is of order $1/n$.  We'll see below that the number of e-folds of inflation in our model is predicted to be $\sim 80$.  In terms of FRW geometry, this is given by
\begin{equation} 80 \sim N_e = \int\ H dt = \int\ \frac{dH}{\epsilon H} \sim - \frac{1}{2\epsilon} {\rm ln}\ (n_{BH} n^3) =  - \frac{2}{\epsilon} {\rm ln}\ (\frac{T_{RH}}{10^{7.5} GeV}) . \end{equation}  In the last two equalities we've approximated $\epsilon$ by a constant, and anticipated the fact that the immediate post inflationary universe is a dilute gas of black holes of radius $n$, with initial co-moving number density $n_{BH} < n^{-3}$ . The universe is reheated by decay of these black holes and its maximal reheat temperature $ \sim 10^{7.5} $ GeV is attained when the initial density of the black holes is maximal.   It's likely that the bound on $n_{BH}$ is stronger than the obvious restriction that the holes be separated by more than their Schwarzschild radii.  At any rate \footnote{In previous discussions of holographic inflation we claimed that the model predicted $r$ of order one.  That claim was incorrect.}, this implies $\epsilon \sim 1/40$   .  As a consequence the tensor to scalar ratio $r$, is nominally $\sim 6 \times 10^{-4}$ . There is an ``order one" relative normalization of the quantum 
fluctuations in black hole mass and angular momentum, which contributes to $r$, and which we do not yet know how to calculate.  Thus, in general we predict 
$ r = s \epsilon^2 $ and the current bound on $r$ gives $\epsilon < \sqrt{\frac{.1}{s}} $.  The relation between $\epsilon$ and the number of e-folds ($ N_e \sim 80$ based on data about the current temperature and the c.c.) is also somewhat flexible and can be modified by a change in the shape of the slow roll trajectory.

The horizon DOF of HST provide an intuitive way of understanding the difference between EFT inflation models and the HST model. Quantum EFT views the entropy of a horizon volume of dS space as the entanglement entropy of that local region, with the rest of the field DOF in space-time, in a unique quantum state, the adiabatic vacuum of linearized metric fluctuations (in co-moving gauge).  This is, implicitly, an enormous fine tuning of initial conditions on modes of the quantized fluctuations which have initial wavelengths much shorter than the Planck scale.  The choice of adiabatic vacuum is usually justified by appeals to the adiabatic theorem, in order to deal with the fact that we don't know the true dynamics of trans-Planckian modes.  That argument {\it does not} remove the fine tuning.  The adiabatic theorem does not apply to generic states of a large quantum system, in which the level spacing is smaller than the natural scale of energy by a factor of $e^{- {\rm Entropy}}$.

In HST, the entropy of a dS horizon is attributed to the fact that the horizon DOF have a fast scrambler Hamiltonian, so that the time averaged density matrix is maximally uncertain.  The dS temperature is a consequence of the fact that particle states are defined by constraints on the horizon DOF, with energy fixed by the constraint.  When erstwhile dS horizon volumes are combined together into the interior of the apparent horizon of a slow roll inflationary cosmology, the time averaged density matrix of individual horizon volumes is still maximally uncertain, but there has been an increase in the total number of DOF.  In order for the fluctuations to be localized, the initial state must be such that the matrices $M(p)$ for the full apparent horizon have the block diagonal form
\begin{equation} \begin{pmatrix} K_1 & 0 & 0 & \ldots \cr 0 & K_2 & 0 & \ldots \cr \vdots & \vdots & \vdots \cr 0 & 0 & \ldots & K_h .\end{pmatrix} \end{equation} $h$ is the number of independent horizon volumes just after inflation ends.  Each of the blocks has the same size, but the action of the Hamiltonian can effectively turn matrix elements on and off, so the properties of these blocks will be time dependent.   If $n$ is the average size of the non-zero blocks, then the time dependent fluctuation will be of order $\frac{\delta n}{n} \sim 1/n$.  In order for the inflationary era to succeed in making localized excitations, the zeroes must persist as the horizon expands, and this is a constraint on the Hamiltonian $H_{in}$ during this era. Vanishing off diagonal matrix elements gives of order $(hn)^2 $ constraints on the horizon DOF, when the horizon radius is $hn$. Note that this is exactly the same counting as the constraint that no more than $h$ black holes of radius $n$ can fit into a horizon of size $hn$, without collapsing into a single large black hole.   This constraint is much less severe than the QUEFT insistence on a unique initial state, but more importantly, {\it it is justified as the most entropic way to make localized excitations}. 

The basic idea of our model is to have $H_{in} (t)$ evolve as in the EHI model, for some finite number, $N_e$ of e-folding times of the asymptotic (asymptotic in the HEI model) dS radius $n$.  The maximal entropy of the causal diamond, from $t = n$ to $ t = N_e n$, is $n^2$ and the time averaged density matrix is maximally uncertain.  The averaging time is $o(n)$.  We will see the necessity for this period of inflation later on.  After time $N_e n$, more matrix elements of $\psi_i^A (p)$ are included in $H_{in} (t > N_e n )$, but they are in a constrained state.  When the horizon size has grown to $k n$, at a time $t_k$\footnote{From this point on, we will take snapshots of the time evolution only at the times $t_k$.  We also ignore trajectories that are more closely spaced than the inflationary horizon radius $n$.} we assume that the state of the system is such that the matrices $M_{nk}  (p,q)$ are block diagonal, with some small blocks of size $n < K_i  < nk - \sum K_i$, and one remaining large block.  The individual blocks are in a typical state of the maximally uncertain ensemble, and evolve on a time scale inversely proportional to the block size.

Our work on scattering in Minkowski space, reviewed in Section 2, shows that as the horizon expands, such block diagonal configurations have many of the qualitative properties of black holes.  They are equilibrium systems which emit particles at the Hawking rate, and interact via a long range $1/r$ potential.  
For any configuration of these black holes, we can solve the HST consistency relations, at least in a coarse grained way, on scales $\gg $ the black hole horizon sizes, by incorporating into the $H_{out} $ of each trajectory, the same state experienced along a trajectory which has some set of black holes in the ${\it in}$ tensor factor of its Hilbert space.  Our knowledge of classical gravitational physics enables us to understand the qualitative evolution of a universe with each of these initial states.  We will first discuss the history of the universe on FRW slices, and then re-interpret it on HST slices.

If the black hole distribution is inhomogeneous, the black holes will collide and form larger black holes.  This lowers the Hawking temperature and makes the black holes more stable.  Such a universe will end up containing a finite number of large black holes.  From the point of view of any given trajectory most of these black holes would disappear into the horizon of dS space in a time scale of order $N$, while those that remained bound to the origin would slowly radiate (both classical gravitational radiation, if the bound configuration contained orbiting black holes, and Hawking radiation) and collapse into a single black hole.  Eventually, that final black hole would evaporate.  At no time in the history of such a universe could any sort of complex organization arise.

Thus, one would expect a configuration of black holes which remained as close to homogeneous as possible.  Exact homogeneity is impossible, because the black holes are finite, and rather small.  Thus, they are chaotic quantum systems with finite entropy. There will be fluctuations of all their properties, including mass and angular momentum, of order
$$\frac{\delta M}{M} \sim \frac{\delta L}{L} \sim S^{-1/2} = n^{-1} ,$$ where $n$ is the Schwarzschild radius in Planck units.  These are fluctuations in the Weyl curvature, which carry spin zero and spin two in Newman-Penrose coordinates.
Note that, since these are fluctuations of collective properties of chaotic macroscopic systems, the decoherence of these fluctuations is obvious.  Quantum interference effects in computations involving these fluctuations are of order $e^{ - n^2}$.  Since, {\it a priori}, $n \gg1$ in the model, and data indicates  that $n \sim 10^5 - 10^6 $, they are entirely negligible.  By contrast, in conventional inflation models, the primordial fluctuations are two point functions of quantum fields in the adiabatic ground state of the slow roll space-time.  The question of why these fluctuations decohere has hardly been touched on in the inflation literature\cite{fstbk}. While we are sympathetic to some of the arguments for decoherence in QFT inflation, the ground states of even large quantum systems can exhibit remarkable amounts of measurable quantum interference.   In the HST model there is no question that we can treat the fluctuations as if they came from a classical probability distribution for small deviations from the background FRW metric.

It is worth dwelling for a moment on the reason for these differences, because it lies at the heart of the distinction between HST and QFT.  In HST, {\it most} of the variables in a causal diamond do not have an approximate bulk QFT description, but are instead a holographically correct reduction of boundary gauge DOF\cite{tbwftoappear}.  Particles are constrained states of these boundary DOF and there are no bulk field DOF.  The fluctuations in the HST model are a consequence of chaotic evolution of a quantum system of entropy $n^2$ among a large class of its states.  In conventional inflation, the fluctuations are quantum fluctuations in a unique ground state, which is by assumption (the adiabatic argument, which justifies using the adiabatic vacuum for erstwhile trans-Planckian modes), separated from any chaotic thermal ensemble by a gap.  

The attribution of horizon entropy to entanglement of two localized regions in space-time in a ``vacuum" state is a common feature of the QFT description of both Hawking radiation and inflation.  In the context of evaporating black holes, it leads to the ``firewall" paradox\cite{ampsetal}.   HST eschews this issue\cite{fw3} by identifying the states responsible for the entropy as real low energy states not captured by bulk QFT.  The analogous states account for the fluctuations we see in the CMB.

So far, nothing we have said connects our picture to the conventional view of inflation.  Along each time-like geodesic in the emergent FRW space-time one sees more and more black holes come into the horizon, as it expands.  The theory predicts a probability distribution for these black holes which is approximately homogeneous, isotropic, and Gaussian.  Homogeneity and isotropy make it easy to satisfy the HST consistency relations between different trajectories, on scales greater than the black hole horizon size $n$.  There are other initial conditions which also satisfy the consistency conditions, which lead to universes dominated by a few large black holes.  
 
The connection with inflation comes, when we compare the view of the universe on HST slices, with that on FRW slices.  Consider a very late HST slice, which intersects an FRW slice whose conformal time is just a bit smaller than $\eta_0$ the position of the pole in $a(\eta )$, which signals the existence of the cosmological horizon.  This trajectory sees black holes come into the horizon, which have been propagated by the Hamiltonian $H_{out} (\eta )$ up until that moment.  The model we are proposing insists that the variables comprising each of these black holes were decoupled from the rest of the system until that HST time slice, which intersects the black hole trajectory, on the FRW time slice $\eta_0 / 2$.  Homogeneity and isotropy tell us that the same must have been the case for our own trajectory, up to that point in conformal time.  
However, a model consisting of a whole set of decoupled systems, each with entropy $n^2$, and the same Hamiltonian, is precisely our description of multiple horizon volumes of the EHI model.   Thus, on FRW slices starting when the horizon size is $n$ and up to conformal time $\eta_0 / 2$, the entire universe underwent inflation.

The number of e-folds of inflation is determined by the future cosmological horizon, saturating the bound of \cite{efoldbound}  .     The observable quantity related to this bound is the range of scales over which we expect to see fluctuations.  In the present model the lower limit of this range is $n \frac{a_{NOW}}{a_I}$, where $a_I$ is the scale factor at the end of inflation, and the upper limit is the cosmological horizon.  We will estimate $N_e$ after we discuss reheating.  First however, we must digress to discuss the symmetry, which determines the spatial distribution of the fluctuations.

\subsection{$SO(1,4)$ Invariance}

We now want to claim that our model guarantees that, as fluctuations enter the horizon of the observer, they approximately obey the constraints of $SO(1,4)$ invariance.  This means both that the Hilbert space carries a unitary representation of $SO(1,4)$, and that the state of the system is invariant.  In order to understand the model we must talk about {\it all} of the DOF in the system, not just those which are within a given trajectory's apparent horizon at a particular time.  Our model assumes that the asymptotic state of the universe is dS space, with Hubble radius $10^{61} L_P$.  As we've argued in the past\cite{dS}, the full set of DOF are a collection of $L N ( N + 1)$ matrices $\psi_i^A (p)$, which are fermionic generators of a super-algebra for each value of $i$ and $A$.  We have 
$ \pi (RM_P)^2 \sim N^2 L$, where $L$ is the number of values of $p$.
The Hamiltonian is written as a function of the matrices \begin{equation} M_i^j (p,q) = \psi_i^A (p) \psi^{\dagger\ j}_A (q) . \end{equation} We will insist on an initial state in which of order $K n N$ matrix elements of all of these matrices vanish, with $K n\ll N$. Furthermore, the state of the non-zero variables is a tensor product of states on which the two non-zero blocks of the matrix operate.   This ansatz is motivated by our description of Minkowski space.  It is the way to incorporate localizable excitations into {\it any} space-time.  
The large $(N - n K)^2 $ block describes states, which reside on the cosmological horizon of dS space.   The $n K \times n K$ block describes the inflationary era and the $p = \rho $ era, which precedes it.   At the beginning of the universe, we  take the sequence of time dependent Hamiltonians for each $n \times n$ block to be those of the $p = \rho \rightarrow p = - \rho$ model described in the previous section.  As described there, the evolution ends with each of these systems thermalized by a Hamiltonian which, if $n \gg 1$, is the $L_0$ generator of an $SL(2)$ algebra.

In order to understand the origin of $SO(1,4)$ we have to understand the action of the $SO(3)$ rotation group of a single trajectory on the entire Hilbert space of the system, including both states that are inside and outside the horizon at any given time.  We will see that $SO(1,4)$ acts only on the Hilbert space of states outside the horizon.  When the horizon has size $nk$, this is spanned by the action of $o([N - n k]^2 )$ variables.   These can be broken up into 
$[(N/n) - k ]^2$ groups of $n^2$ variables.  The Hamiltonian $H_{out} (k)$ has the form 
$$\sum_{l=1}^{[(N/n) - k ]^2} L_0 [l] ,$$ where the sum is over the independent groups. $L_0 [l]$ is the Cartan generator of the approximate $SL(2)$ algebra, which we identified in the $p = \rho$ phase of cosmological evolution.  

The full Hilbert space of the theory is spanned by the action of the $N(N + 1) L$ operators $\psi_i^A (p)$ and their adjoints.  It is acted on by $SU(2)$ in the obvious way.   The $\psi_i^A$ are a finite basis in the space of measures on the spinor bundle of the sphere.  They include all measures that can be written as a finite sum of spinor spherical harmonics, with maximal angular momentum $N - 1/2$.  Consider a collection of $o(n^2)$ linearly independent measures, maximally localized in a region of area $n^2$ surrounding some point $\Omega (l) $ on the sphere.  Each such measure is constructed by taking some matrix $f_{iB}^{jA} (l) $, of rank $n^2 (n + 1)^2$,  and writing $(\psi_f )_i^A = \psi_j^B f_{iB}^{jA} $. Now act on those measures with rotations, localizing them at another point $l^{\prime}$ a distance of order $n$ away.   We insist that the matrices satisfy $$ f(l) f(l^{\prime} ) =  f(l^{\prime} ) f(l) = 0 .$$  In this way, we decompose our degrees of freedom into operators localized in regions of fixed area on the sphere.  There are of order $(N/n)^2$ independent sets of angularly localized operators.  To visualize this, we can take a spherical grid, obtained by triangulating the faces of an icosahedron into $\frac{1}{20} (N/n)^2$ simplices, and set each $\Omega (l) $ to the center of one of the triangles.  

Now we introduce the following approximate $SO(1,4)$ generators

\begin{equation} J_{04} = \sum_l L_0 (\Omega (l) ) ,\end{equation}
\begin{equation} J_{\pm i} = \sum_l [ \Omega_i (l) L_{\pm} (\Omega (l) ) ] . \end{equation}
$J_{ij}$ are the generators of $SU(2)$, acting on the spinor variables. The commutation relations of these generators fail to be those of $SO(1,4)$ in two ways.  The $SL(2)$ commutators between $L_0$ and $L_{pm}$ have corrections of order $1/n$.  In addition, the commutators between the $J_{ij}$ 
and the other generators give the correct scalar and vector transformation laws only in the limit $N \rightarrow \infty$ in which the set of $\Omega (l)$ becomes dense on the sphere.  Note that giving the $\Omega_i (l)$ an icosahedrally symmetric distribution on the sphere is enough to guarantee the correct commutators for the generators $J_{i\ \pm}$, up to the $1/n$ corrections to the $SL(2)$ algebra.  

The scrambling time evolution generated by $H_{out} = J_{04}$ in individual blocks of $n^2$ variables is enough to guarantee that the density matrix, averaged over a time of order $n {\rm ln}\ n$, commutes with generators $J_{04}$ and $J_{\pm i}$.  However, we must impose unnaturally constrained initial conditions on the system, which guarantee that at the time a given block of $n^2$ matrix elements moved from $H_{out}$ to $H_{in}$, all of the variables that connect it to the rest of the matrix are set equal to zero.  This is our {\it topik\`{e}sthropic} constraint.  The question arises of how we choose the constrained variables.

It is important to realize that there are an infinite number of ways to embed $SU(2)$ into the group of all unitary transformations $\psi \rightarrow V \psi W $.  Make any choice, and conjugate it by an element of this $U(N) \times U(N+1)$ group.  The Hamiltonian $H_{in} (t_k)$ is invariant under a $U(kn) \times U(kn + 1)$ subgroup of this group, while $H_{out} (t_k)$ is invariant under permutations of our chosen blocks of $o(n^2)$ variables.  These transformations do not, in general, respect the assignments we have made of blocks to points on the sphere.  They correspond to measure preserving transformations, which are not in general differentiable or even continuous.
Thus, the only thing in the dynamics that picks out a particular way of choosing the $SU(2)$ group is the interpretation of the state inside the horizon in terms of localizable objects.  

We have already discussed the constraints which follow from the fact that the distribution of black holes inside the horizon must be approximately homogeneous and isotropic.  Our theory is quantum mechanical, so what this means is that the density matrix, which gives the probability of finding a black hole in some angular region on the sky at some time, commutes with the $SU(2)$ group that we identify with rotations in space.  We have many unitarily equivalent $SU(2)$ algebras on the full Hilbert space, under each of which the variables $\psi_i^A (p)$ transform like the $N \otimes N + 1$ representation.  The choice of a sequence of variables $\psi_i^A (p)$ which are in the $h \times h +1$ representation when the horizon size is $h$, picks out a particular $SU(2)$ as $h \rightarrow N$.  We want the density matrix to be invariant under this $SU(2)$.   Recall that we defined the breakup of the DOF outside the horizon into multiple blocks of size $n^2$, by interpreting one such block as localized within a certain solid angle, and the rest as $SU(2)$ rotations of that block.  This $SU(2)$ should now be identified with the one picked out by dynamics inside the horizon ({\it i.e.} by the fact that the increase in horizon size is achieved by adding angular momentum multiplets to the system).  It's then clear that the Hamiltonian $J_{04}$ commutes with rotations in the limit that there are many blocks.  

The only further thing that we should point out is that this does not require a significant increase in the fine tuning of initial conditions.  All configurations with the same total number, $N_{BH}$ of black holes inside the ultimate dS horizon have roughly the same entropy deficit $N_{BH} n N$ for their initial conditions.  The only exceptions are those with a very small number of black holes.  Initial conditions of this type lead to a universe where any given trajectory is in causal contact with a few black holes and their decay products.  They are small fluctuations of the EHI model (with Hubble radius $N$), and have no complex local excitations.  

The local $SL(2)$ symmetry of individual black holes, combines with the emergent $SU(2)$ we have discussed, to fill out the full $SO(1,4)$ group, and we predict scale invariant fluctuations (apart from the explicit slow roll factor in the relation between the gauge invariant measure of scalar fluctuations, $\zeta$, and the scalar curvature fluctuation).  {\it In particular, we predict the tensor tilt of primordial fluctuations to be zero.}  In the next section we'll see that Hawking decay of the black holes gives rise to another distribution of random gravitational radiation, whose spatial profile is tied to that of the scalar fluctuations.  It's amplitude is suppressed by a factor $1/g$, relative to the scalar 
two point fluctuation, where $g$ is the multiplicity of particles with mass small compared to the Hawking temperature $ \sim 10^{-6} M_P$.  Depending on the value of the slow roll parameter, this distribution might dominate over the primordial one.  Its tilt is the same as that of the scalar fluctuations.

Although we are currently describing the Hamiltonian along a given time-like trajectory, we can now associate each of the black holes with different time-like trajectories as well.  In our description of Minkowski space, we explained how copying the dynamics of one trajectory's ``in" Hilbert space, into the ``out" Hilbert space of another trajectory, is the way to satisfy the HST consistency conditions in situations where there are localized excitations.  The different systems of $n^2$ DOF evolve under the action of
their own copy of $L_0$ for varying amounts of time.  This is because we are measuring time as proper time along a given trajectory, and choosing the time slicing of the causal diamond to be such that at each instant, a given space point is described as it appears to a detector traveling along that trajectory.  These are the hyperbolic time slices of Fig. 1, and {\it not} the FRW slices. 

If we take another copy of the same quantum system, and assign it to a different time-like trajectory, the sequence of black holes it encounters will be different, as will the time slices on which the quantum mechanics is defined.  We can read off the intersection of the causal diamonds of the two trajectories from the background geometry.  We've chosen our distribution of black holes to be homogeneous and isotropic, and each is associated with a subset of the variables.  This makes it easy to verify that the coarse grained consistency conditions are satisfied.  We do not yet have a construction satisfying the consistency conditions for Planck spaced trajectories.  Note that homogeneity and isotropy of the multi-trajectory construction is consistent with $SO(1,4)$ invariance.  Within the Hilbert space of a single trajectory, the $SO(1,4)$ transformations predict that physics will look the same at other points in the flat spatial slicing of dS space, and that is indeed what we find by assigning the same system to each point on the lattice of trajectories.

Note that exponentially expanding scale factors and Hamiltonians for fields in the background space-time appear nowhere in the HST description.  The causal structure of an inflationary universe ${\it does}$ appear explicitly, and we use that to define inflation.  The real difference between the HST and field theory descriptions is that the fluctuations are statistical fluctuations of the quantum state of large chaotic quantum systems, rather than quantum fluctuations in an adiabatic ground state.  This is an initial condition which is much less tuned than
the assumption of a unique ground state, but more importantly, within the HST formalism this tuning is required in order to get a cosmology with local excitations.

The resulting cosmology is more or less fixed in terms of a small number of parameters: the slow roll parameter $\epsilon$ (really a slowly varying function of time rather than a single parameter), the inflationary Hubble scale $n$, in Planck units, the black hole density at the end of inflation, $n_{BH}$ and the Hubble scale of the final c.c., $N$.  These are related by inequalities
$$ 1 \ll n \ll N ,$$ $$n_{BH} < n^{-3} .$$  They determine the reheat temperature of the universe, and roughly the number of e-folds.  The observable quantity, for which the number of e-folds is a proxy in conventional inflation models, is the range of scales over which we expect to see scale invariant fluctuations. In our model that range is 
\begin{equation} e^{N_e} = \frac{N a_I}{n a_{now} } . \end{equation}  $a_I$ is the value of the scale factor at the end of inflation, while $n$ and $N$ are the Hubble scales during inflation and at the end of the universe.  We will estimate this quantity after explaining the reheating mechanism in the HST model of inflation.

\subsection{Reheating}

In the previous section we've argued that, apart from the intrinsic fluctuations in black hole mass and angular momentum, the black hole distribution is homogeneous and isotropic.  However, fluctuations grow in a matter dominated universe, which is certainly the correct characterization of the immediate post-inflationary phase of the HST cosmology.  In order to distinguish it from the matter dominated phase that recently ended, we call this the Dilute Black Hole Gas (DBHG) era of the HST universe.

Let $n_{BH}$ be the number of black holes per unit comoving volume, on an FRW slice immediately after inflation ends.  We define the scale factor to be $1$ at this time.  If $Q = (n \epsilon )^{-1}$ is the strength of primordial perturbations, then the fluctuations become of order one when the scale factor is $a = n\epsilon$, so that the average distance between black holes is \begin{equation} d_{BH} = n_{BH}^{ - 1/3} n\epsilon .\end{equation} Note that this occurs at a time \begin{equation} t_{fluct} =  (n \epsilon)^{3/2} \ll n^3. \end{equation}   Thus, fluctuations become $o(1)$ long before the black holes can decay.
If these fluctuations have time to collapse to form larger black holes, then the universe will become dominated by a collection of large, almost stable black holes.  A long time later, these will decay into the dS horizon.  Although such a universe has a period of local physics, it does not allow for any kind of complex organization to occur.  Thus, we must choose $n_{BH}$ in such a way that typical black holes evaporate before they can combine and collapse.  

It turns out that as long as we satisfy the obvious bound $n_{BH} < n^{-3}$, which says that the distance between black holes is larger than their Schwarzschild radius, then, taking into account the matter dominated expansion of the universe, the in-fall time for an order one fluctuation to accumulate more black holes around it, and coalesce with them, is longer than the evaporation time $n^3$.  So most of the black holes will decay, though we may expect a few exceptional regions with large meta-stable black holes, because the time scale for growth of fluctuations is so much shorter than the evaporation time.  It's amusing to speculate that these might be the seeds that are needed in the modern theory of galaxy formation, but we have not investigated this in detail.

The reheat temperature is roughly just the fourth root of black hole energy density $n n_{BH}$, redshifted by a factor $t_{decay}^{-2} \sim n^{-6}$.  The maximal reheat temperature is thus $o(n^{-2})$.  If $g$ is the effective number of massless species at the time of black hole decay, then the black hole decay time is shorter by a factor of $g^{-1}$ while the formula relating black hole energy density to the reheat temperature becomes $n n_{BH} t_{decay}^{-2} = g T_{RH}^4 $.  So the maximal reheat temperature is of order $g^{1/4} n^{-2}$ and the temperature scales like $(n_{BH} n^3)^{1/4} $ for smaller initial densities.  It's likely that this factor must be less than one.

The size of primordial fluctuations is $\sim 10^{-5}$ observationally, and scales like $(n \epsilon)^{ - 1}$ in the HST model. Combining this with the rumors from the Planck and BICEP2 collaborations suggesting a firm observational bound on the tensor scalar ratio \begin{equation} r = s \epsilon^2 < 0.1 , \end{equation} gives an upper bound on the reheat temperature of 
\begin{equation} T_{RH}  = g^{1/4} 10^{-10} {\epsilon^2} < g^{1/4} 10^{-11} / s . \end{equation}  The parameter $s$ has not yet been calculated in HST.  Recall that in conventional inflation models, $r = 16 \epsilon$, so we might err in assuming that $ s = o(1)$. Assuming it is $o(1)$, and that $g \sim 10^3$ we get a maximal reheat temperature of order $6 \times 10^8 (n_{BH} n^3)^{1/4}$ GeV.  Thus the primordial black hole density can be as much as $10^{40}$ times lower than its theoretical maximum, before we run into trouble with Big Bang Nucleosynthesis.
Depending on $n_{BH}$, this cosmology could be compatible with a variety of mechanisms for baryogenesis.  In the conclusions we will discuss the possibility of asymmetries generated in the decay of primordial black holes.

We can use these considerations to estimate the range of scales over which our model predicts primordial fluctuations. As noted in the previous section this is
\begin{equation} e^{N_e} = \frac{N a_I}{n a_{now}} .  \end{equation}  The first factor, $N/n$ is $ N Q \sqrt{\epsilon} = 10^{56} \sqrt{\epsilon}$.  $Q$ is the size of primordial fluctuations. The second factor is

\begin{equation} \frac{a_I}{a_{RH}} \frac{T_{now}}{T_{RH}} = t_{decay}^{2/3}  g^{-1/4} n^{2} (n_{BH} n^3)^{-1/4} (.25 \times 10^{-31}) \end{equation} \begin{equation} = n^4 (g n_{BH} n^3)^{-1/4} (.25 \times 10^{-31}) = Q^{-2} \epsilon^{-1} (g n_{BH} n^3)^{- 1/4} (.25 \times 10^{-31}) . \end{equation}
If we use $g \sim 10^3$ we get 
\begin{equation} e^{N_e} = e^{ 2.3 (34.25) - 1.39 - .25 {\rm ln}\ (n_BH n^3 + \epsilon^2) } . \end{equation}  This gives
\begin{equation} N_e = 78.78 - .25 {\rm ln}\ (n_BH n^3) - .5 {\rm ln}\ ( \epsilon) > 79 .\end{equation}
Assuming $\epsilon$ is approximately constant in time, we get $\epsilon \sim 1/40$, which implies $r = .0006 s$.  This is certainly compatible with the rumors from the Planck and BICEP2 collaborations suggesting that we will soon have a firm observational bound of about  $r < 0.1$, unless $s$ is very large.  

\subsection{Comparison With Data}

A detailed comparison of the HST model with data will be presented in \cite{tbwftoappear}.  Here we give a brief summary of the results.  In short, the fit of the scalar two point function to the data succeeds for the same reason it succeeds in slow roll inflation.  Our model gives an approximately scale invariant spectrum, modified by a prefactor which is a function of time.  The prefactor in the HST model is different from that in field theory models.  It comes only from the slow roll factor relating the gauge invariant $\zeta$ to scalar curvature fluctuations in co-moving gauge.  In the field theory models, the scalar curvature fluctuations are those of a scalar field in the adiabatic vacuum of the slow roll geometry, while in HST they are exactly scale invariant.  We've noted above that the curvature fluctuations in field theory are down by a factor of $\sqrt{\epsilon}$ relative to those in HST.  At the moment, the HST models have another free parameter, characterizing the choice of complementary series representation for the quantum operator .   These are the representations determined by applying the dS/CFT\footnote{We emphasize that we are {\it not} doing dS/CFT, but merely using the reader's familiarity with that literature to identify a class of unitary representations of $SO(1,4)$.} dictionary to a bulk massive field in the regime where the associated boundary dimension is real.  The field theory analysis uses the singular limit of these representations corresponding to vanishing bulk mass.  We hope that further analysis of the approximate $SO(1,4)$ invariant quantum mechanics of the matrix model will pin down the value of this parameter.  It should also help to calculate the coefficient $s$ in the relation $r = s \epsilon^2$.  Despite all these differences, there is no way to distinguish between the two kinds of models on purely observational grounds, solely from the scalar two point function.  Both models have the free choice of the background FRW scale factor $a(t)$.

There has been a lot of discussion in the literature about constraints on the effective field theory of inflation, coming from the fact that, for single field, slow roll inflation, the data seems to indicate super-Planckian field excursions, anathema to hard core effective field theorists.  It has long been clear to string theorists, that the effective field theories of string compactifications {\it always} involve super-Planckian field ranges, and that those regimes are controlled by weakly coupled approximations to quantum gravity.  The real problem in those regimes, is to find a potential capable of sustaining inflation, which `` could be computed reliably in string theory"\cite{steinhardt}\cite{banksdine}.  The most significant constraints come from the considerations of \cite{bdfg}\cite{amnv}, but the authors of those papers never claimed more than parametric control over such calculations.  No forces can be tuned to be weaker than gravity by an arbitrarily large factor.  No potential coming from a valid theory of quantum gravity can vary over a range larger than the Planck scale by an arbitrarily large factor. Claiming anything more is tantamount to claiming understanding of the corrections to Hawking's calculation of black hole decay and entropy for an arbitrary model of quantum gravity.   The data on CMB fluctuations require at most a range of field values of order $\sim 30 m_P$, and we have no valid reason to reject such potentials.  Indeed, there are myriad examples in the literature, which give evidence for the existence of moderately super-Planckian ranges of variation in ``controlled" string theory calculations.

In HST, the FRW model during inflation is a Jacobsonian, rather than a Wilsonian description of the underlying quantum model of cosmology.  There is no real ``inflaton field" or potential and no constraint on $a(t)$ other than the weakest energy constraint on the Einstein tensor, which guarantees that the laws of thermodynamics are not violated.  From our point of view, it would be nice if string theorists could show that constraints on the low energy field theory describing fluctuations are Minkowski or AdS space, forbade potentials that could describe the data on the CMB, but we believe that the arguments of the previous paragraph preclude that. 

As we've emphasized, the dominance of scalar over tensor two point functions follows from cosmological perturbation theory applied to a slow roll geometry with scalar and tensor co-moving curvature fluctuations of the same order of magnitude.  The main observational difference between the two models at the two point function level is that field theory predicts a tensor tilt of $r/8$, while HST predicts vanishing tensor tilt.  The rumors from the Planck and BICEP2 collaborations suggest that we will soon have a firm observational bound on $r$\footnote{Note that the theoretical bound $r < .0006 s$ in the HST model, is not relevant to this comparison.} makes these predictions almost indistinguishable over the range of scales available from the CMB, large scale structure, and simulations of galaxy formation.  Perhaps the PIXIE mission\cite{pixie} to search for short wavelength gravitational radiation could sort them out.

The situation with respect to the gravitational wave two point function is actually a bit more complicated, because gravitational waves are also present in the Hawking radiation of decaying black holes.  A fraction $\sim 2/g$ of the energy of the decay goes into gravitational waves.  With $g \sim 10^3$ this might compete with the primordial signal if $\epsilon$ is very small.  The spatial distribution of this radiation will follow that of the black holes and so the two point correlation will share the scalar tilt of $n_S = .96$, which is very different from the standard inflationary prediction.  Our current uncertainty about the size of $\epsilon$ required to fit the scalar data and the non-observation of tensor fluctuations, prevents us from saying which of these two contributions dominates.

Finally, venturing into the realm of science fiction, we point out that field theoretic and HST models of inflation make very different predictions for the tensor three point function\cite{holoinflation2}. The assumption of approximate $SO(1,4)$ invariance allows for three different form factors, but field theory models predict that one of the three dominates by a factor of $(H/m_P) \sim 10^{-6}$, while the parity violating form factor vanishes to all orders in the effective field theory expansion.  HST predicts that the two parity conserving form factors are of comparable magnitude, and we have not yet found an argument that the parity violating contribution vanishes.  It might be that there are different HST models, which either have or do not have a reflection symmetry that forbids the parity violating 3-point function.  This deserves further study, but the sad truth is that the prospect of actually measuring these fluctuations within the lifetime of the current generation of younger physicists, is dim.

Indeed, the correct conclusion from our analysis of the data on cosmological fluctuations is that the general framework of classical cosmological perturbation theory and approximate $SO(1,4)$ invariance can account for all of the data gathered so far, without committing to a specific model.  Models whose foundations are radically different, can give indistinguishable predictions for extant data, and may not be distinguishable for a long long time.  Details of the HST predictions will appear in \cite{tbwftoappear}.

\subsection{The Matrix Model and the Dilute Black Hole Gas.}

We have described the matter and radiation dominated eras, which follow inflation in our model, in terms of the semi-classical physics of black holes.
While we know that the Matrix model gets qualitative properties of black hole physics right in Minkowski space, it is worth emphasizing that it can also account for the coarse grained geometry in a more direct manner.  Our model takes snapshots of the sub-luminal time evolution of the universe at times $t_k$ characterized by an apparent horizon radius $k n$.  The time $t_k$ coincides with FRW time at the position of the trajectory, and follows the hyperbolic slices of Fig. 1 away from the trajectory. The key to the derivation of the conventional FRW equations is that the evolution is designed to be {\it isentropic}.   The entropy of the initial state is set by the the parameters $n_{BH}, n$ and $N$, combined with the conditions of approximate homogeneity and isotropy, and it does not change until the black holes decay.  $n_{BH}$ controls the rate at which black holes enter the horizon.  

The fast scrambler Hamiltonian ${\rm Tr}\ P(M_{kn})$, must have an overall scale $1/ (kn)^2$ in order to obtain the proper $1/r$ scaling for the large impact parameter scattering of black holes\cite{holonewton}.  On the other hand, this gives a scrambling time $ t_s = kn\ {\rm ln} (k) $, which must be equal to the proper time in the diamond in order for localized objects to retain their distinct identity.  We conclude that, since the horizon radius is linear in $t$,  $a(t)$ must have the single component FRW form
$a \sim t^{\frac{2}{3(1 + w)} }$, up to small corrections.  The bulk calculation of the number of black holes in a horizon volume then gives
$$ N_{BH} \sim n_{BH} (kn)^{\frac{1 + 3w}{(1 + w)} } .$$ On the other hand, from the matrix model construction of black hole states as block diagonal matrices we see that $N_{BH} < kn $, so the only allowed equation of state with non-negative $w$, is $w =0 $.  An equation of state with negative $w$ would correspond to a matrix model in which the number of black holes did not grow linearly with $kn$ .  This is of course possible with $w = -1$, and corresponds to the models we have discussed where the black hole distribution is non-uniform.
We don't know how to construct a matrix model consistent with any other negative $w$ equation of state.  For a homogeneous isotropic distribution of black holes, the matrix model predicts a pressureless equation of state.

\subsection{Environmental Selection of Baryogenesis and Dark Matter from Black Hole Decay?}

Black hole decay in an expanding universe is an irreversible process, which violates baryon number and CP (if there are any CP violating processes in nature, they operate near a black hole horizon).  On the other hand, the usual thermal spectrum of Hawking radiation does not exhibit any baryon asymmetry.
This is an artifact of the thermodynamic approximation.  The correct statement is that black hole decay is described by a probability distribution for baryon number emitted during the decay, which is symmetric around zero and peaked there.  For a black hole of radius $n$ in Planck units, with a Hawking temperature much higher than the mass of the lightest particle carrying baryon number, any particle emission event is likely to carry some baryon number, and the system does a random walk on the baryon number axis every time interval of order $n$.  The black hole decays away in $n^2$ such intervals, leaving over a residual baryon number $\Delta B$ with probability $P(\Delta B)$ .  $P$ is probably well approximated by a Gaussian centered at zero with a width $< (\Delta B) \sim \frac{1}{n} .$   The probability for a give baryon to entropy density of the universe $\epsilon_B$ is obtained by treating the Dilute Black Hole Gas as a collection of independent, identical Gaussian random processes, and this gives $< \epsilon_B^2 >  \sim (n_{BH} )^{- 1/2}\times  \frac{1}{n^2}$ .

It's clear that the most probable universe produced at the beginning of the radiation dominated era has $\epsilon_B = 0$, but this is not necessarily the case.  Anthropic pressures might dictate a  non-zero value.  The analysis of this is quite complicated.  On the one hand, for a wide range of $n_{BH}$ the reheat temperature is high enough to accommodate a variety of well known models of baryogenesis, though we don't know if any of those models come naturally out of HST.  On the other hand, the only known models of TeV scale physics compatible with the HST hypothesis of Cosmological SUSY Breaking, have a dark matter candidate which is a baryon-like object of a new strong gauge group\cite{tbetal}, whose density is determined by a primordial asymmetry.

Furthermore, this {\it pyrma-baryon} number current  $J^{\mu}_{PB}$ is coupled to the baryon and lepton densities of the standard model by dimension six current current couplings of roughly calculable magnitude.  A primordial asymmetry in one of these quantum numbers acts as a chemical potential for the other, and can generate an asymmetry for it during a period of thermal equilibrium.  As a consequence the question of asymmetries gets mixed up with a host of other anthropic questions about HST cosmology.

As we have described it, the model has, in addition to any asymmetries, the parameters $n, n_{BH}. \Lambda, $ and the slow roll FRW $H(t)$.  $\Lambda$ is simply fixed, while the others are plausibly choices of initial conditions.
There is a multi-verse model of HST cosmology in which asymptotically dS cosmologies with a collection of values of $\Lambda$ are embedded in a background $p = \rho$ universe by using the Israel junction conditions to match the dS horizon to a marginally trapped surface in the cosmological space-time.
This model allows for anthropic selection of both $\Lambda$ and the initial conditions.  It would also allow dS bubbles with different primordial asymmetries.  The anthropic constraints are complicated and involve all of the parameters.

For example, Weinberg's galaxy formation bound reads
$$\Lambda < c \rho_D Q^3 , $$ where $c$ is a constant and $Q$ the amplitude of primordial scalar fluctuations.  $\rho_D$ is the density of dark matter at the beginning of the matter dominated era.  In the HST model, this depends both on the primordial asymmetry in the dark matter and the reheat temperature.  Both the reheat temperature and $Q$ depend on $n$.  The baryon asymmetry, which is also subject to anthropic pressure, depends on the dark matter asymmetry.
Finally, $\Lambda$ itself affects low energy particle physics (and through it, nuclear physics and chemistry) because of its' connection to SUSY breaking.
We have not yet sorted through this complicated tangle, but it is certainly within the realm of possibility that the anthropic constraints will give an adequate explanation of why our universe obtained a non-zero asymmetry in some approximate global quantum number despite the improbability of non-zero asymmetry in the ensemble of all possible worlds.

It's also possible \cite{hook} that CP violating effects involving both the black hole and FRW metrics, could generate baryon and/or pyrma-baryon asymmetries directly.

\section{Conclusions}

We can summarize our model of inflation in a way which makes only minimal reference to the explicit matrix models we've proposed, but does follow the tenets of HST.  We assume that the final state of a quantum model of a future asymptotically dS space has a finite dimensional Hilbert space and a Hamiltonian which is a fast scrambler.  We take the further hint from the form of Schwarzschild-dS entropy, that empty dS space is the maximally uncertain density matrix and localized states are constrained states of the system with entropy deficit $2 \pi R E$, for energy $E$ as measured by a static observer.  The most entropic way to have localized energy $E$ is to make a black hole of that energy.  More generally, in {\it any} causal diamond in any Lorentzian space-time, the Unruh effect is explained by saying that the degrees of freedom in that diamond live on the boundary of the diamond, and are thermalized by a fast scrambling Hamiltonian whose maximal eigenvalue is of order $1/R$. Localized objects are low entropy constrained states of these DOF with an entropy deficit of order $ER$.  $E$ is the quantum number that counts the total energy of the localized objects, in the limit $R/L_P \rightarrow\infty$, in which $E$ becomes conserved.  The time evolution described by this Hamiltonian is proper time evolution along the geodesic connecting the tips of the causal diamond.  Accelerated trajectories merely experience a slowing down of proper time (redshift of $E$) for localized objects bound to the trajectory, and so become more sensitive to the slow dynamics on the horizon.

Cosmology is an evolution of the system in which the Hamiltonian is time dependent and splits into $H_{in} (t_k) + H_{out} (t_k)$, where $H_{in} (t_k)$ is constructed from operators which operate in a tensor factor of the Hilbert space that has entropy $\pi R(t_k )^2$ , where $R(t_k)$ is the size of the apparent horizon in Planck units.  $H_{out}$ is constructed from operators that commute with all of the variables in that tensor factor.  We choose the Hamiltonian $H_{in} (t_k)$ to be the fast scrambling Hamiltonian described above, with number of DOF and operator bound on the time dependent Hamiltonian described by $R(t_k)$.  $R$ can only change by integer values, and this corresponds to a Planck scale discretization of time.  Over longer time scales, $R(t)$ becomes a smooth function and defines the cosmological evolution.  As stated, we concentrate on models where $R( \infty ) = R \gg L_P$.   In the very early universe we always assume the state of the Hilbert space on which $H_{in} (t)$ acts is generic\footnote{The Hilbert space has a time dependent tensor factorization into {\it in} and {\it out} spaces.  The linearity of quantum evolution allows us to always follow the evolution of factorized states at each time, even though the initial conditions or the consistency conditions of HST, may force entanglement between the in and out factors of the actual time dependent state.}.  We also assume that for $t \gg L_P$ in the very early universe, the Hamiltonian becomes that of a $1 +1$ CFT.
We've seen that this leads to a description of the evolution of a generic state as a cosmology with scale factor $a(t) = \sinh^{1/3} (3t/R) $.  This behavior is independent of our choice of $H_{out}$, and it describes a universe that has no localized excitations besides those which arise as random thermal fluctuations during the de Sitter period.  

In order to have localized excitations we {\it must } choose a low entropy state.  The maximal entropy localized state for a given entropy deficit in the full Hilbert space, is a state consisting of black holes.  We can now use our intuition about black holes on the FRW slices of cosmology.  If the black hole distribution is non-uniform, two mechanisms act to prevent the formation of more complex structures in the universe.  First, black holes form bound states, and coalesce to form larger black holes, which are more stable. Furthermore, black holes that are not bound together will be driven outside each other's horizon in a time of order $R$.   Black hole decay in such a universe will mostly consist of a slow dribbling of radiation, followed by a single explosion.   There is no mechanism for forming long lived structures with complex patterns of energy transfer.

This indicates that despite being disfavored by the entropy of the initial state, situations with a near homogeneous gas of black holes are favored by a very weak form of a bio-thropic principle.  Note that the entropy deficit of any initial state has a large term $E R$, which is the same for homogeneous and inhomogeneous distributions of black holes, with the same total energy.  The minimal amount of in-homogeneity we can expect comes from statistical fluctuations of the black hole state.  They are finite chaotic quantum systems.  The mass and angular momentum of the black holes are thermodynamic variables and thus have fluctuations of order $1/\sqrt{S}$ as the black hole wanders chaotically among the different states in the thermal ensemble.  Each black hole will enter the horizon at a different point in this chaotic cycle, since the horizon evolution $R(t)$ is in no way synchronized with this microscopic dynamics.  The fluctuations will show up as fluctuations in the local Weyl curvatures. These are clearly decoherent fluctuations whose statistics obey the rules of classical probability with accuracy $e^{-S}$.  

The connection between this picture and the theory of inflation comes when we examine the consistency conditions on $H_{out}$ imposed by the model of $H_{in} $, which uncovers a near homogeneous distribution of black holes as the horizon expands.  The definition of a localized black hole is a quantum system
which is {\it not} interacting with the horizon degrees of freedom.   Here we must invoke the matrix model, to understand the nature of the potential interaction.
$H_{in} (t)$ is $\frac{1}{R^2 (t)} \times $ the trace of a polynomial in the $R(t) \times R(t) $ matrices $M_{R(t)}$.  A black hole is a generic state of a set of fermions $\psi_i^A (p)$ whose matrices are $n \times n$, with $n \ll R(t)$.  In order for such a block to be non-interacting, we must be in a state where the off diagonal matrix elements between the small block and the large $R(t) - n$ block, vanish.  To insure that we have prepared such a state, we choose $H_{out} (t)$ to be a sum of non-interacting Hamiltonians for $n \times n$ matrix blocks, each of which is the approximate $L_0$ generator of the $SL(2)$ algebra we invoked to obtain the scaling laws of the $p= \rho \rightarrow dS$ universe, with dS radius $n$.  With this choice of $H_{out}$ the system is prepared in a way that allows localized black holes to appear in the Hilbert space of $H_{in}$, as the horizon expands.  We add a collection of $2h + 2$ of these independent subsystems to the {\it in} Hilbert space, when the horizon expands from $h n$ to $(h + 1) n$.  Some of these are added to the horizon block, and some become localized black hole blocks, with the split determined by the parameter $n_{BH}$.  We saw that consistency between the bulk and matrix model pictures forces us to choose the equation of state $p = 0$ for the FRW metric that gives a coarse grained description of the process.

This was the description of the entrance of black holes into the horizon of a particular trajectory, from the point of view of that trajectory.
However, we must also consider consistency with the description of a time-like trajectory, which goes through the space-time position of the black hole at the time that it enters the horizon of the first trajectory.   The observer along that trajectory is experiencing inflation, because its {\it in} Hilbert space has maintained constant entropy $n^2$ for proper times between $n$ and the spatial distance, in inflationary Hubble units at which the black hole crosses the horizon .  Since the universe is homogeneous on FRW slices, our original trajectory must also experience inflation up till that FRW time. 

The variables $\psi_i^A$ for the full dS space-time, with $N \gg n \gg 1$, carry a representation of $SU(2)$.  Using it, we localized individual subsets of variables, each with $o(n^2)$ DOF at $o(N/n)^2$ angles $\Omega_i$ on the sphere, and combined $SU(2)$ and the $SL(2)$ algebras of individual blocks of DOF to construct generators an approximate $SO(1,4)$, which controls the behavior of low order correlation functions of fluctuations.

The matrix model we've presented, gives a complete mathematical formulation of a system whose qualitative behavior was described in the last few paragraphs.  Indeed, it gives many such formulations, depending on the choice of the commutation relations of the fermions and the polynomials $P (M)$.  We believe that, among all these models, only a very limited set will give completely consistent theories of quantum gravity, when we impose the constraints that come from requiring that we obtain consistent descriptions of overlapping causal diamonds for trajectories that are in motion relative to each other.  In this paper, we've only described trajectories that are at rest in the same frame.

To conclude, we'd like to summarize both the major differences between the model of this paper, and our previous work, and to compare our model to standard QUEFT inflation.  The present model is based on our understanding of locality in asymptotically flat space, whereas our previous work tried to construct an explicit lattice field theory.  There, the approximate $SO(1,4)$ symmetry, which explains the shape of scalar fluctuations, was postulated, whereas here it arises naturally from the combination of the $SL(2)$ invariance of the $p = \rho$ model and the $SU(2)$ that is built into the definition of our variables.  We also understand it to act only on the part of the Hilbert space that is outside the horizon.  

This connection between inflation outside the horizon (which implies via the HST consistency conditions that inflation occurred also in the past of a given trajectory) and the appearance of a roughly homogeneous dilute black hole gas inside the horizon, is also new.  This goes along with the new realization that once physics is local, we must differentiate between the equal time surfaces of the HST model, and the FRW surfaces.  The picture of the post inflationary universe as a dilute black hole gas harks back to some of our earliest work on inflationary cosmology, gives us a rather constrained model of reheating, and explains the physical origin of the scalar and tensor fluctuations as mass and angular momentum fluctuations of the black holes, which are viewed as finite chaotic quantum systems.  This should lead to a determination of the absolute relative normalization of the two kinds of fluctuation.

Apart from the fact that HST is a well defined quantum mechanical model, its main advantage over QUEFT models of inflation is that its choice of initial conditions is much less fine tuned, and that the fine tuning is justified as the minimal amount necessary in order to get local physics in an asymptotically dS universe.  More precisely, the adjective minimal applies to models in which there are only a few localized black holes inside the cosmological horizon, but we feel that it's pretty obvious why a more homogeneous distribution of black holes is necessary for the existence of complex organization.  

Another advantage of HST over QUEFT models is that the model is much more highly constrained, and depends on a very small number of parameters.  All HST models satisfy the approximate $SO(1,4)$ invariance, which we demonstrated in \cite{holoinflation2} is sufficient to explain all extant cosmological data, when combined with Maldacena's general theorems about soft scalar fluctuations.  There is a wide variety of QUEFT models, which can explain all sorts of peculiar forms for non-Gaussian fluctuations.  While many models are ruled out by current data, the number of surviving models is large.
We believe that it will be very difficult to falsify the general idea of QUEFT inflation without measuring the tensor three point functions.

Finally let us note that the decoherence of the fluctuations and their interpretation as classical statistical fluctuations of the Weyl tensor follow from the black hole interpretation in HST: black holes are large chaotic quantum systems and their mass and angular momentum are collective coordinates.  In QUEFT models, the fluctuations originate as quantum fluctuations in a unique state, separated, according to the adiabatic argument, from the rest of the Hilbert space by a gap large compared to the Hubble frequency.  One needs special arguments to justify their decoherence.

\section{Appendix: Scattering of Massless Particles and the Super-BMS Algebra}

Let us make the convention that outgoing momenta carry positive energy.  The GSBMS generators are operator valued {\it half measures}\footnote{Mathematicians, missing the chance at a silly joke, call these {\it half densities}.  A half measure is a quantity whose square is a measure. More generally, the product of two different half densities is a measure.} on the momentum space light cone $P^2 = 0$.  That light cone is a dual representation of the conformal boundary of Minkowski space, with each momentum in the forward (backward) light cone representing the outgoing (incoming) momentum through a point on future (past) null infinity.  The singular joining point $P = 0$ is dual to space-like infinity, and it is important to keep the angular dependence of the zero momentum point. Boundary conditions at $P=0$ are conditions at space-like infinity.

The past and future GSBMS algebras are 
\begin{equation} [ \psi^{\mp}_{alpha} (P,p) , \psi^{\mp}_{\dot{\beta}} (Q,q) ]_+ = \mp \delta (P\cdot Q) M_{\mu} (P,Q) \sigma^{\mu}_{\alpha\dot{\beta}} Z(p,q) .  
\end{equation}
Other anti-commutators vanish.  The delta function ensures that $P$ and $Q$ are collinear when the anti-commutator is non-vanishing.  $M(P,Q)$ is the momentum that is closest to zero in absolute value\footnote{There is a mathematical formulation of these algebraic relations in which the algebra of functions on half of the light cone is realized as the  tensor product of functions on the sphere with the sub-algebra of commuting projections in the hyperfinite $II_{\infty}$ factor\cite{connes,tb11}.}  Each momentum $ P_{\mu} = P (1, {\bf \Omega})$ is associated with a unit 3 vector ${\bf \Omega}$ and a projector having ${\rm Tr}\ e_P = P $.  Then the vector $M_{\mu} (P,Q) = {\rm Tr}\ e_P e_Q (1, {\bf \Omega})$.  .   The labels $p,q$ are other quantum numbers, which may flow out to infinity.  They can be thought of as labeling the eigen-sections of the spinor bundle over a compact manifold, with a cutoff on the spectrum of the Dirac equation on that manifold\cite{tbjk}.   In the language of string theory, we may view the bosonic generators $Z(p,q)$ as wrapped brane charges on the internal manifold.  

Scattering representations of the GSBMS are those for which the smeared generators $\psi (f) \equiv \int \psi_{alpha} (P,p) f^{\alpha)}(P) $ vanish on all states, with the following exceptions 
\begin{itemize} 
\item The support of $f$ at non-zero $P$ vanishes outside of a finite number of non-overlapping spherical caps with finite opening angle.  
\item The support of $f$ at $P \rightarrow 0$ vanishes in annuli surrounding those caps.
\end{itemize}
The zero momentum generators represent soft particles which are ``almost gauge modes", and whose emission and absorption amplitudes are fixed by gauge invariance\cite{weinbergstrom}.  In taking a limit from finite causal diamonds, they represent momentum flows carrying momentum that goes to zero through any finite area on the holographic screen at infinity, but with integrated total momentum different from zero. The basic idea of these constraints is that all soft modes, which are emitted or absorbed almost co-linearly with 
finite momentum particles, are incorporated into ``Sterman-Weinberg jets"\cite{stermanweinberg}.  The amplitudes for a collection of incoming jets to transform into a collection of outgoing jets are expected to be infrared finite, as long as their opening angles are finite.  Note that the considerations leading to the careful treatment of zero momentum modes at null infinity are important in all dimensions.  In dimension $d = 4$ amplitudes vanish unless a finite total amount of energy is emitted into these `` zero momentum modes". In higher dimensions amplitudes with `` zero momentum emission" must be included to get an exactly unitary S matrix.

The GSBMS identifies the limit of zero opening angle jets as multiplets of supersymmetric massless particles, if there is a zero mode of the internal Dirac operator for which $Z(0,0) = 1$.  Indeed, in that case, the algebra multiplying the collinear delta function is just that of a single super-particle of momentum $P_{\mu}$.  Particles defined in this way are automatically identical, essentially for the same reason that they are identical in QFT.  Particles are excitations of a field on the holographic screen at the conformal boundary of Minkowski space, and permutation of particles does not change the field configuration.  The connection between spin and statistics follows from the fact that the algebra defining the asymptotic Hilbert space is graded, and the half integral spin generators are charged under a local $Z_2$ gauge symmetry acting on the null cone.

Note that the GSBMS algebra is compatible with the condition
\begin{equation} P_{\mu} \bar{\sigma}_{\dot{\beta}\alpha}^{\mu} \psi_{\alpha} (P) = 0 . \end{equation}  This condition says that the fermionic variables at each point on the null cone are null plane spinors of the local momentum, or, what is the same thing, that they are sections of the spinor bundle over the 2-sphere at infinity. This is the key connection to the variables $\psi_i^A (p)$ of this paper.

The S-matrix maps a copy of the GSBMS algebra in the past\footnote{Actually, the the past and future algebras differ by a minus sign because incoming momenta have negative time components and linear combinations of the LHS of the algebra are manifestly positive.} into a future copy
\begin{equation} \psi_{\alpha}^+ [f] = S^{\dagger} \psi_{\alpha}^- [f] S .\end{equation}




\end{document}